\documentclass[12pt,preprint]{aastex}
\shorttitle{RGBbump}
\shortauthors{Nataf et al.}
\usepackage{natbib}
\usepackage{float}
\usepackage{array,longtable}

\title{OGLE-III Detection of the Anomalous Galactic Bulge Red Giant Branch Bump: Evidence of Enhanced Helium Enrichment}
\author{D. M. Nataf\altaffilmark{1}, A.Udalski\altaffilmark{2,3}, A.  Gould\altaffilmark{1}, M.H. Pinsonneault\altaffilmark{1}}
\altaffiltext{1}{Department of Astronomy, Ohio State University, 140 W. 18th Ave., Columbus, OH 43210}
\altaffiltext{3}{Optical Gravitational Lens Experiment (OGLE)}
\altaffiltext{2}{Warsaw University Observatory, Al. Ujazdowskie 4, 00-478 Warszawa,Poland}
\email{nataf@astronomy.ohio-state.edu}

\begin{document}

\begin{abstract}
We measure the red giant branch bump (RGBB) of the Galactic bulge, the most metal-rich RGBB ever detected. The RGBB luminosity function peaks at the expected brightness, but its number density is very low relative to Galactic globular cluster calibrations, implying the Galactic bulge has a higher helium enrichment parameter  ${\Delta}Y/{\Delta}Z \geq 4.0$ for $Y \sim 0.35$  rather than the standard $2.0$ with $Y=0.27$, which we suggest may be a common feature of galactic spheroids. The RGBB is $(0.71 \pm 0.02)$ mag fainter than the red clump (RC) in $I$ toward the densest stellar regions imaged by the OGLE-III Galactic bulge photometric survey, $(|l| \leq 4, 2 \lesssim |b| \leq 4)$. The number density of RGBB stars is $(12.7 \pm 2.0)$ \% that of RC stars. The brightness dispersion of the RGBB is significantly lower than that of the RC, a result that is difficult to explain as the RGBB luminosity is known to significantly vary with metallicity. Sightlines that have two RCs have two RGBBs with similar properties to one another, an expected outcome if the Milky Way's bulge is X-shaped. We also find preliminary evidence of the Galactic bulge asymptotic giant branch bump, at a brightness of $\sim$1.06 mag brighter than the RC in $I$ and with a number density $\sim$1.5\% that of the RC. Accounting for the RGBB has a small effect on the best-fit parameters of the RC, shifting its best-fit peak brightness and reducing its brightness dispersion by $\sim$0.015 mag each.
\end{abstract}

\section{Introduction}
\label{sec:Introduction}
The red giant branch bump (RGBB) and the red clump (RC) are two observables of the color-magnitude diagrams (CMDs) of low and intermediate mass metal-rich populations which correspond to specific post main-sequence phases of stellar evolution. The RGBB occurs first, during ascent of the red giant (RG) branch. As the hydrogen burning shell expands, it eventually comes into contact with the convective envelope, which provides additional fuel. The star then gets fainter until the extra fuel is consumed whence it begins brightening again. As the star crosses the same luminosity region three times, there is an increase in the number density of stars at that luminosity \citep{1997MNRAS.285..593C}. The RC occurs after the degenerate helium core has accreted enough mass for core helium burning induced by the helium flash, it is the horizontal branch of an old, metal-rich population \citep{2001MNRAS.323..109G}. The relative number counts and brightness for these two populations are a function of their composition. In particular, the RGBB gets fainter relative to the RC as metallicity increases \citep{1997MNRAS.285..593C,1999ApJ...518L..49Z,2003A&A...410..553R,2010ApJ...712..527D}, and the lifetime of the RGBB decreases with increasing helium abundance \citep{2001ApJ...546L.109B,2010ApJ...712..527D}, while that of the RC increases \citep{1994A&A...285L...5R}.

Whereas the RC has had a vibrant history as a tracer of Galactic bulge reddening \citep{1996ApJ...460L..37S,2003ApJ...590..284U,2004MNRAS.349..193S,2009ApJ...696.1407N}, structure \citep{1997ApJ...477..163S,2005MNRAS.358.1309B,2007MNRAS.378.1064R,2008A&A...491..781C,2010ApJ...721L..28N,2010ApJ...724.1491M} and dynamics \citep{2007MNRAS.378.1165R,2010A&A...519A..77B}, there has been little study of the Galactic bulge RGBB. Indeed, we find only three brief mentions. \citet{2003A&A...399..931Z} state that the RGBB is expected to be 0.7 mag fainter than the RC in $J$, and that the two will be vertically mixed on a CMD. The RGBB was also discussed in two recent investigations of the double RC toward sightlines near the bulge minor axis and at least 5 deg removed from the plane \citep{2010ApJ...721L..28N,2010ApJ...724.1491M}. Both works commented on a small, secondary excess of stars fainter than the RC, which they suggested may be due to the RGBB without further analysis.

Interestingly, \citet{2010ApJ...724.1491M} argued that the double RC may have been observed before but not recognized as such. In their Figures 4 and 5, \citet{2005MNRAS.358.1309B} plot a $(J-K,K)$ CMD of a field at $(l,b)=(0,1)$ for which there is a second overdensity $\sim$0.7 mag fainter than the RC. This second peak also shows up in an analysis by \citet{2005ApJ...621L.105N} of photometry along the strip $(|l|<10, b = +1)$. We argue the secondary density was in fact the RGBB and not the double RC. Within our own data, we find that the RGBB signal is stronger for sightlines closer to the plane, an effect that can be explained by the lower geometric (and thus brightness) dispersion for those sightlines. 

\begin{figure}[H]
\includegraphics[totalheight=0.74\textheight]{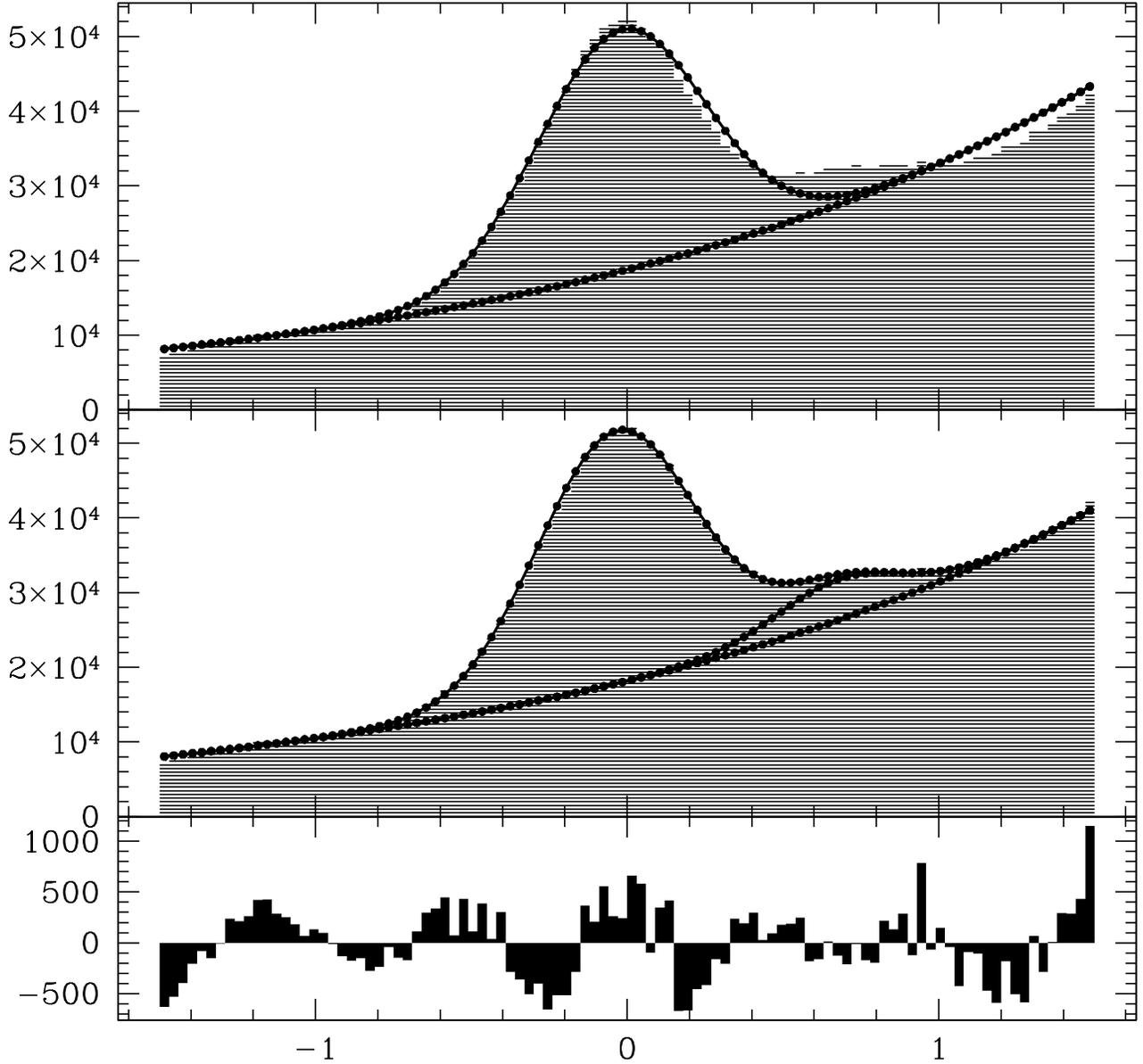}
\caption{The magnitude distribution for 2,836,575 OGLE-III point sources that are within 4 deg$^{2}$ of $(l,b)=(0,0)$, that have an $I$ measurement within 1.5 mag of the local RC centroid, and that are no more than 0.35 mag bluer than the nearest RC. The top panel shows the fit to a Gaussian plus a power-law for the RC and RG sources, the middle panel shows the fit once a second Gaussian for the RGBB is added. ${\Delta}{\chi}^2 = 6670$.  The bottom panel shows the residual to the latter fit. Prominent and significant are a few non-Gaussianities to the RC and an excess of stars $\sim$1.1 mag brighter than the RC, likely due to the AGBB. }
\label{Fig:markovscriptPaper}
\end{figure}

In this Letter, we use the OGLE-III Galactic bulge $VI$ photometry dataset and find that the RGBB is $\sim$0.71 mag fainter than the RC in $I$, and that its number density is $(12.7 \pm 2.0)$\% that of the RC.  Accounting for the RGBB slightly affects the parameters of the RC, the number density of RC stars shrinks by $\sim$2\%, the peak brightness shifts by $\sim$0.015 mag toward brighter mag, and the brightness dispersion decreases by $\sim$0.015 mag. The fit and distribution is shown in Figure \ref{Fig:markovscriptPaper}. We may also have detected the asymptotic giant branch bump (AGBB), at  $\sim$1.1 mag brighter than the RC with a number density of order $\sim$1.5\% that of the RC. Sightlines with two RCs have two RGBBs with similar properties. In section \ref{sec:Data}, we describe the OGLE-III observations. We summarize the evidence for the Galactic bulge RGBB in section \ref{sec:Evidence}. We present the method of the clump-centric CMD in section \ref{sec:ClumpCentric}, and our method of measurement in section \ref{sec:Fitting}. Each of the measured RGBB parameters are presented and interpreted in section \ref{sec:Parameters}. We discuss our findings as well as the prospects and directions of future research in section \ref{sec:Discussion}. 

\begin{figure}[H]
\includegraphics[totalheight=0.46\textheight]{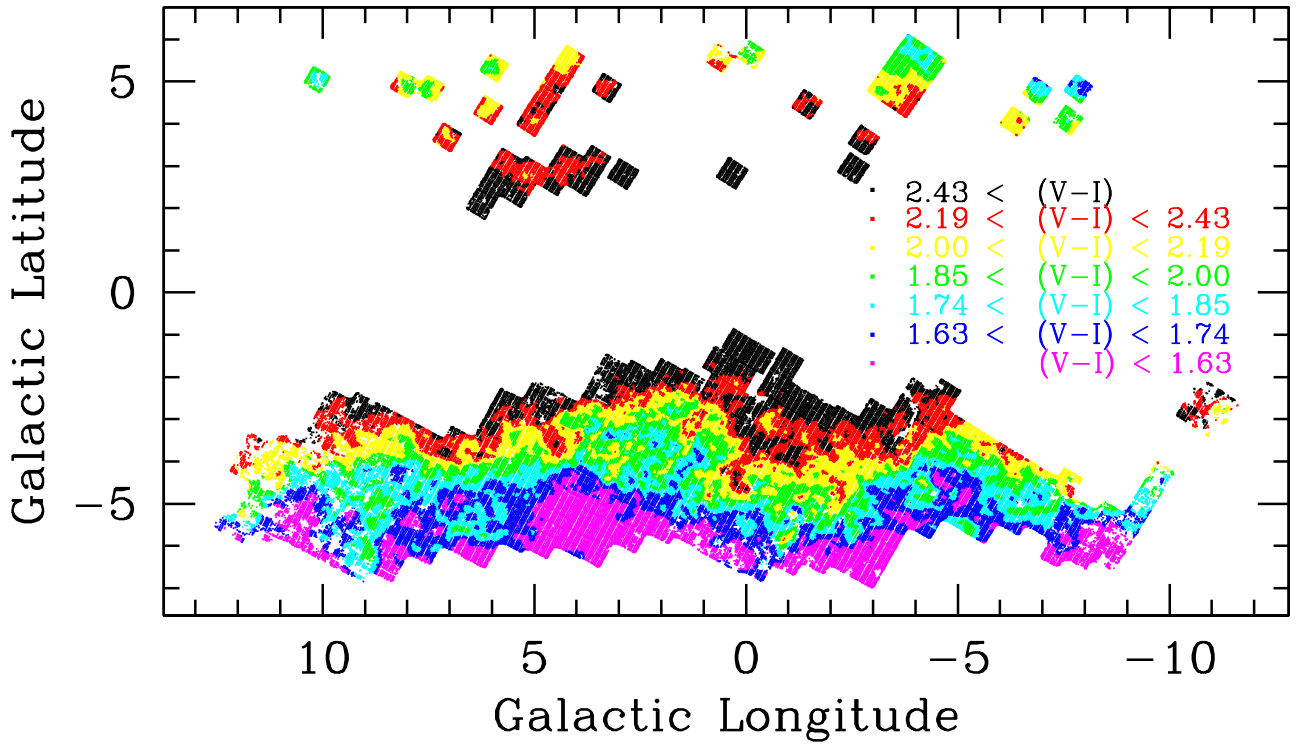}
\caption{The OGLE-III Galactic bulge observing coverage, color-coded by the $(V-I)$ color of the RC. The RC gets redder closer to the plane due to higher interstellar extinction and reddening -- the intrinsic color of the Galactic bulge RC is $(V-I) \sim$1.08 \citep{2010A&A...512A..41B}. As the RGBB is $\sim$0.71 mag fainter than the RC, and OGLE-III is complete down to 21st magnitude, we are able to detect the RGBB over nearly the entire field of view.}
\label{Fig:JustReddeningMap}
\end{figure}

\section{Data}
\label{sec:Data}
OGLE-III observations were obtained from the 1.3 meter Warsaw Telescope, located at the Las Campanas Observatory in Chile. The camera has eight 2048x4096 detectors, with a combined field of view of $0.6^{\circ}\times
0.6^{\circ}$ yielding a scale of approximately 0.26$\arcsec$/pixel. We use observations from a mosaic of 267 fields directed toward the Galactic Bulge, which are almost entirely within the range $0<|l|<10$ and $2<|b|<7$. The photometric coverage is shown in Figure \ref{Fig:JustReddeningMap}. 
Of the 267 fields, 37 are at northern latitudes. In this study we do not investigate the fields BLG200, 201, 202 and 203, located in and around $(l,b)=(-11,-2.5)$, due to the anomalous CMDs toward those sightlines, possibly due to a smaller scale differential reddening. More detailed descriptions of the instrumentation, photometric reductions and astrometric calibrations are available in \citet{2003AcA....53..291U} and \citet{2008AcA....58...69U}.

Our quantitative results are obtained from 2,836,575 OGLE-III point sources that are in subfields centered within 4 deg$^{2}$ of $(l,b)=(0,0)$. We chose this subsample as it is more robust -- it has a higher density of stars, lower disk contamination, lower geometric dispersion, no double clump, and no controversy of multiple Galactic bars implying distinct stellar populations. Our error estimates are based on comparisons with other coordinate selection criteria. 

\section{Three Lines of Evidence Toward the Red Giant Branch Bump}
\label{sec:Evidence}
There are three lines of evidence that lead us to the conclusion that the RGBB is a measurable feature of Galactic bulge $(I,V-I)$ CMDs. It appears as a statistically significant secondary peak in our magnitude histograms, there are two such additional peaks for sightlines with two RCs, and the peak is approximately where it is expected to be from globular cluster calibrations. 

The first can be recognized from the distribution of differences between the observed distribution of RC+RG stars and their model fit of a power-law plus a Gaussian. There is a clear excess above the fit for stars 0.5-1.0 mag fainter than the RC. The absolute area of the residuals is reduced by up to 75\% once a second Gaussian is added to account for the RGBB. ${\chi}^2_{DoF}$ for the fit is 75.38 without an RGBB. When a free Gaussian for the RGBB are added, ${\chi}^2_{DoF}$ drops to 5.34 - an unambiguous statistical signal. We note that the large remaining value of ${\chi}^2_{DoF}$ indicates that 8 free parameters are not sufficient to fully describe the data. That is not surprising as a variable slope to the red giant branch, differential reddening on small scales, foreground disk contamination, an asymptotic giant branch bump, and non-Gaussianities in the RC and RGBB are all legitimate possibilities for additional parameters.

\begin{figure}[H]
  \begin{center}
\includegraphics[totalheight=0.48\textheight]{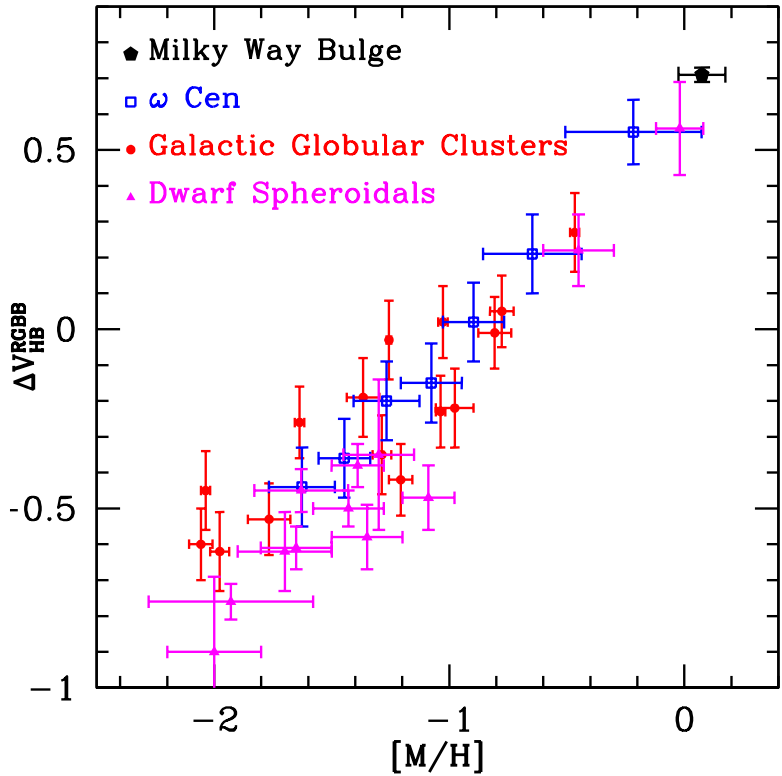}
\end{center}
\caption{The ${\Delta}V_{ZAHB}^{RGBB}$ for the 15 Galactic globular clusters analyzed by \citet{2010ApJ...712..527D} are shown as red circles, the 7 RGBBs of the cluster $\omega$ $Cen$ \citet{2004AJ....127..958R} are shown as empty blue squares, and the 12 RGBB measurements toward 9 local dwarf spheroidals are shown as filled magenta triangles (Monachesi et al. 2011, Monelli et al. 2010, and references therein).} 
\label{Fig:BumpEmpiricalHistory}
\end{figure}

\begin{figure}[H]
  \begin{center}
\includegraphics[totalheight=0.7\textheight]{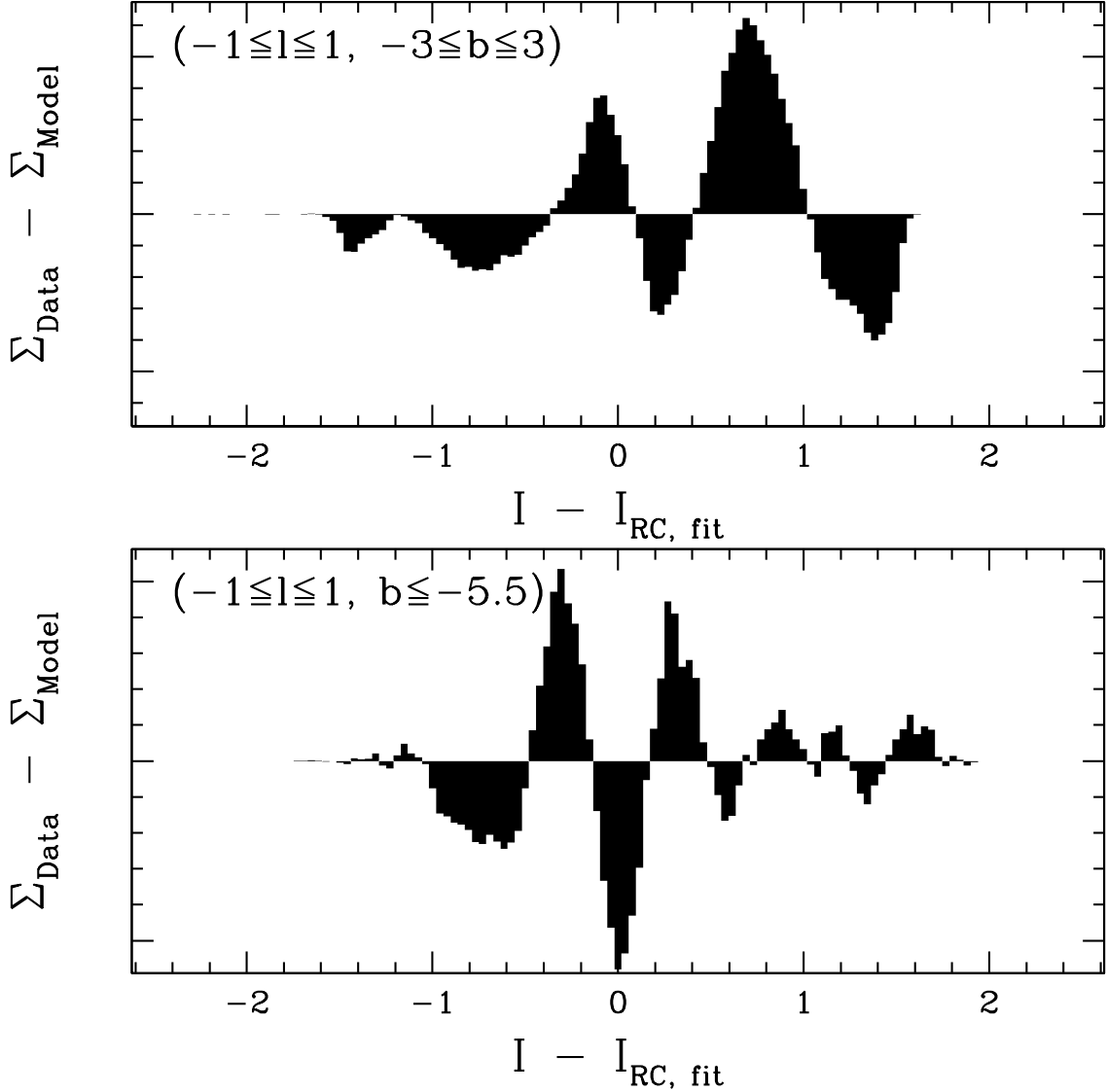}
\end{center}
\caption{The sum of a residuals to a single-clump+power-law fit over thousands of sightlines containing no more than $\sim$1200 RC+RG stars to two general areas the bulge, $(|l|{\leq}1,|b|{\leq}3)$ in the top panel, and $(|l|{\leq}1,-8{\leq}b{\leq}-5.5)$ in the bottom panel. The residuals in the top panel, toward single-clump areas, are dominated by the RGBB and the requirement of normalization. Those toward the bottom panel reveal the presence of two RCs, as well as two RGBBs that are approximately equally well-populated and equally well-separated to their respective RCs.} 
\label{Fig:DoubleBump}
\end{figure}

The second line of evidence is related to the double-RC of the Galactic bulge, seen on sightlines within $\sim$2 degrees of the bulge minor axis and at least $\sim$5 degrees removed from the plane. The two RCs are approximately equally populated, have equal or very nearly equal $(V-I)$ and $(J-K)$ color, and are separated in brightness by $\sim$0.5 mag \citep{2010ApJ...721L..28N,2010ApJ...724.1491M}. For double-clump fields, combining data from different sightlines with different reddening is non-trivial, however we can use the residuals to a single-clump fit on small scales as a preliminary diagnostic. The signal is clear: two clumps and two bumps. If the two clump-bump pairs are found to have the same separation in brightness, and the same population fraction, it would rule out explanations of the double clump based on age and chemistry and confirm an X-shaped Milky Way bulge as the explanation. Both those statements appear approximately correct based on the distribution of residuals. If detailed analysis confirm that the two populations have the same metallicity as one another, they will also show whether they have the same metallicity as the bulge giants toward sightlines closer to the plane, thereby providing an independent test and calibration of the metallicity gradient detected by \citet{2008A&A...486..177Z}. Unfortunately, the method used in most of this paper cannot be directly applied to double-clump fields, due to the varying separation between the two RCs and their varying relative population densities, as well as the fact these variations are not yet well-parametrized. In Figure \ref{Fig:DoubleBump} we plot the sum of residuals to the fits of single RC + RG distribution to thousands of sightlines toward both single-clump fields and double-clump fields. The residuals for the double-clump fields betray not only the presence of a second RC, but also of two RGBBs. 

The third line of evidence is that the brightness peak of the RGBB we report here, ${\Delta}I^{RGBB}_{RC}=0.71$, is consistent with that expected from stellar evolution models and globular cluster calibrations. In their analysis of 54 Galactic globular clusters observed with HST, \citet{2003A&A...410..553R} reported a difference in brightness ${\Delta}F555W$ between the RGBB and the Zero Age Horizontal Branch (ZAHB) evaluated at the RR Lyrae instability strip. They found that ${\Delta}F555W=0.45$ for NGC5927 and $0.53$ for NGC6624, two metal-rich globular clusters  \citep{2009A&A...508..695C}. For the metallicity of the Galactic bulge, [M/H] $\approx 0.0$ \citep{2006ApJ...636..821F, 2007ApJ...661.1152F, 2008A&A...486..177Z, 2010A&A...512A..41B}, the best-fit model presented by \citet{2003A&A...410..553R} has ${\Delta}F555W=0.639$ mag for a 10 Gyr stellar population, and ${\Delta}F555W=0.741$ mag for a 12 Gyr stellar population. Similar results are obtained by \citet{2010ApJ...712..527D} in an investigation of ${\Delta}V_{ZAHB}^{RGBB}$ for 15 Galactic globular clusters. We plot various examples from the literature as well as our own in Figure \ref{Fig:BumpEmpiricalHistory}.

\section{A New Diagnostic for Stellar Populations: The Clump-Centric Color-Magnitude Diagram}
\label{sec:ClumpCentric}
We introduce the clump-centric color-magnitude diagram (CCCMD) as a diagnostic means to study the fine details of large stellar populations in the presence of differential reddening and geometry. It is the color and magnitude of every star within a large angular range relative to the nearest RC centroid.

CMDs are among the most powerful observational tools in Astronomy. Unfortunately, optimizing their construction for stellar populations such as that of the Galactic bulge is a non-trivial task. Over small angular scales, their diagnostic power will be limited by Poisson noise, restricting the analysis of a stellar population to only its most prominent features. For example, within the CMDs used in our previous work on the double RC \citep{2010ApJ...721L..28N}, we used CMDs just large enough to contain 1200 RC+RG stars, so as to minimize any possible gradients. With that number we would only expect $\sim$30 RGBB stars per sightline, rendering it a difficult population to identify let alone classify as it is mixed with the RC and RG populations.

Over large angular scales, differential reddening and geometry both become  problems. \citet{2004MNRAS.349..193S} found that the optical reddening $E(V-I)$ can vary by over $\sim$1 mag within 1 degree. Worse, the extent of differential reddening is larger for sightlines closer to the plane, which have a higher density of stars. This would produce a diagonal smearing over any large CMD. This smearing would have no clear form as both the slope of the reddening law, $R_{I} = A_{I}/E(V-I)$ and the total reddening have patchy and discontinuous variations. Dereddening is typically used as a solution to this problem. In their study of the Galactic bar, \citet{2007MNRAS.378.1064R} dereddened each field to the intrinsic color of the RC. An issue with this method is that the slope of the reddening law is known to vary with direction even though it is not known precisely how it varies with direction. A change in $R_{I}$ of $\sim$0.1 over 2 magnitudes of reddening leads to an error of 0.2 mag in the dereddened brightness. 

Further, it is also known that within the coordinate range $(-4 \leq l \leq 4)$ the brightness of Galactic bulge RC stars varies by $\sim 0.4$ mag due to the orientation of the Galactic bar \citep{1997ApJ...477..163S,2005MNRAS.358.1309B,2007MNRAS.378.1064R}. This effect is in some ways more fundamental than the first effect, as it cannot be evaded by using ``reddening-free'' near-IR or mid-IR data. The obvious countermeasure to assume a Galactic bar model would be an imperfect solution. The actual offset due to the bar is unknown at the level of $\sim$0.1 mag. In addition, the brightness variation with longitude is known not to be independent of latitude. \citet{2005ApJ...621L.105N} found that the variation of brightness with longitude for sightlines toward $b = 1$ is shallower than that found further from the plane, an effect that could be due to a additional inner bar.  

These two factors can be almost completely eliminated by using a CCCMD. By constructing a grid of $\sim$30,000 RC $(V-I,I)$ measurements across the OGLE-III bulge sky, we achieve no worse than $4.5'$ resolution. Over that range, the variation in geometry is negligible, and the effect of differential reddening is minimized to its expected extent over that length scale - typically a few  hundredths of a magnitude or less. By amassing a larger number of stars, we can make detailed measurements of structures such as the RGBB and possibly the asymptotic giant branch bump (AGBB), as well as investigate possible systematics in the RC parameters. 

\section{Fitting for the Red Giant Branch Bump}
\label{sec:Fitting}
We fit for the properties of the RGBB in a two-step process. We first fit for the RC without accounting for the RGBB in a manner described in \citet{2010ApJ...721L..28N}. The CCCMD is produced by subtracting from each star the $(V-I,I)$ color and magnitude of the nearest measured RC. Stars that are nearest to the RCs located in zones of rapid differential reddening, low stellar density, or $V_{RC} \geq 19.5$ are left out of the CCCMD. 

We then fit for the combined RC+RG+RGBB population using the clump-centric color-magnitude cut
\begin{eqnarray}
  -0.35 < (V-I)_{CC} \nonumber \\
 -1.5 < I_{CC}  < 1.5,
\label{EQ:constraint2}
\end{eqnarray}
and the model
\begin{equation}
N(m) = A\exp\biggl[B(I-I_{RC})\biggl] + \frac{N_{RC}}{\sqrt{2\pi}\sigma_{RC}}\exp \biggl[{-\frac{(I-I_{RC})^2}{2\sigma_{RC}^2}}\biggl]+\frac{N_{RGBB}}{\sqrt{2\pi}\sigma_{RGBB}}\exp \biggl[{-\frac{(I-I_{RGBB})^2}{2\sigma_{RGBB}^2}}\biggl]
\end{equation}
where $A$, $B$, $N_{RC}$, $I_{RC}$, and ${\sigma}_{RC}$ are as described in \citet{2010ApJ...721L..28N}; and $N_{RGBB}$, $I_{RGBB}$, and ${\sigma}_{RGBB}$ are the analogous parameters for the RGBB. 

\section{Four Measurable Parameters for the RGBB}
\label{sec:Parameters}

The fitting routine of section \ref{sec:Fitting} leads to four measurable parameters for the RGBB:
\begin{equation}
  {\Delta}I^{RGBB}_{RC} = I_{RGBB} - I_{RC} 
\end{equation}
\begin{equation}
  f_{Bump} = \frac{N_{RGBB}}{N_{RC}} 
\end{equation}
\begin{equation}
  EW_{RGBB} = \frac{N_{RGBB}}{A\exp\biggl[B({\Delta}I^{RGBB}_{RC})\biggl]}
\end{equation}
\begin{equation}
  {\Delta}{\sigma}^2 = {\sigma}_{RGBB}^{2} - {\sigma}_{RC}^{2} 
\end{equation}

\subsection{The Brightness Parameter ${\Delta}I^{RGBB}_{RC}$}
The first parameter, ${\Delta}I^{RGBB}_{RC}$, is the difference in peak brightness between the RGBB and RC, and is $0.71 \pm 0.02$ mag. This is consistent with the expectation from globular clusters. We use $I$ rather than $V$ because the RC has a thinner Gaussian in $I$, allowing a clearer separation of the two populations. This is expected. In the models of \citet{2001MNRAS.323..109G}, $M_{V,RC}$ for an old population has variations with metallicity $\sim$2$\times$ that of $M_{I,RC}$, an effect confirmed in the empirical investigation of \citet{2010AJ....140.1038P}, who looked at the RC of 15 nearby galaxies observed with HST. Moreover, any residual differential reddening will be $\sim$2$\times$ as significant in $V$. These two effects render the bulge RC non-horizontal in $V$, further complicating the fitting routine.

There is a difference between this parameter and that predominantly used in the literature. We computed ${\Delta}I^{RGBB}_{RC}$ whereas most results present ${\Delta}V^{RGBB}_{ZAHB}$. However, these two values should be very nearly equal as the two largest biases are not large and go in opposing directions. Firstly, ${\Delta}V^{RGBB}_{RC}$ should be a little larger than ${\Delta}I^{RGBB}_{RC}$ because the RC will be a little bluer than the RG stars, the bias is expected from stellar theory but is consistent with negligible in our data. Conversely, ${\Delta}V^{RGBB}_{ZAHB}$ will be a little smaller than ${\Delta}V^{RGBB}_{RC}$ as the ZAHB is the dimmest phase of horizontal branch evolution, however in our analysis of globular cluster data we find that this effect can be no more than $\sim$0.05 mag. We thus adopt the approximation ${\Delta}I^{RGBB}_{RC} = {\Delta}V^{RGBB}_{ZAHB}$. 

It is reassuring that the Galactic bulge has the faintest RGBB relative to its horizontal branch as it is the most metal-rich RGBB detected thus far. Galactic globular clusters do not typically reach metallicities as high as [M/H] $\approx 0.0$, and those that do have substantial differential reddening \citep{2001A&A...376..878O} or multiple stellar populations \citep{2009Natur.462..483F}, effects that render the RGBB harder to detect. This expansion of the parameter space at the metal rich end follows recent, complementary detections of the RGBB in younger and more metal-poor  systems, those of the nearby dwarf galaxies. There have been detections of the RGBB toward the Sculptor dwarf spheroidal galaxy \citep{1999ApJ...520L..33M}, Ursa Minor \citep{2002AJ....124.3222B}, Sagittarius \citep{2002ApJ...578L..47M}, and Sextans \citep{2003AJ....126.2840L}, and M32 \citep{2011ApJ...727...55M}. \citet{2010ApJ...718..707M} recently reported on the detection of the RGBB toward Cetus, IC1613, LGS 3 and Tucana. The observed relation between ${\Delta}V^{RGBB}_{ZAHB}$ and [M/H] is shown in Figure \ref{Fig:BumpEmpiricalHistory}, with the \citet{1984ApJS...55...45Z} metallicity scale assumed for the Galactic globular clusters and $\omega$ $Cen$.

We estimate [M/H] via the conversion function suggested by \citet{1993ApJ...414..580S}:
\begin{equation}
\rm{[M/H]} = \rm{[Fe/H]} + \log(0.638*10^{[\alpha/\rm{Fe}]} + 0.362)
\end{equation}
and an [$\alpha$/Fe]$=+0.4$ for the Galactic globular clusters and $\omega$ $Cen$, and [$\alpha$/Fe]$=+0.25$ for the Galactic bulge and M32.

\subsection{The Population Parameters $f_{Bump}$ and $EW_{RGBB}$}
The total number density of Galactic bulge RGBB stars is down by a factor of $\sim$2-3 from that expected from globular cluster calibrations, either relative to the RC or to the RG branch. For both parameters, the Galactic bulge stellar population is a lot more similar to that of M32 \citep{2011ApJ...727...55M} than to that of Galactic globular clusters at comparable metallicity, which suggests a common factor in the stellar evolution of galactic spheroids.

We find that the ratio of the number of RGBB stars to RC stars  is $f = (12.7 \pm 2.0)$\%, and that the equivalent width of the RGBB $EW_{RGBB} = (0.104 \pm 0.020)$ mag. Both of them parametrize the total population of the RGBB, but the former does so by comparing its numbers to that of the RC, and the latter does so relative to that of the underlying RG branch. Both have their relative strengths, $f$ is less likely to be biased by disk-contamination, whereas $EW_{RGBB}$ may be easier to interpret theoretically as it compares two populations before the helium flash. We do not use the standard strength calibrator $R_{Bump}$ \citep{2001ApJ...546L.109B} as it would be extremely vulnerable to disk contamination and photometric incompleteness for stars fainter than $V \approx 21$. 

As these are both new parameters previously undefined in the literature, we had to compute their value in globular clusters. We first used a subsample of globular clusters observed with the Hubble Space Telescope \citep{2002A&A...391..945P}, that have an updated metallicity \citep{2009A&A...508..695C}, and are densely-populated. We correct for their photometric completeness functions. This gave us the value of $EW_{RGBB}$ for 30 globular clusters. The mean and sample deviation are 0.32 and 0.18 mag, and there is a statistically significant correlation with metallicity, $EW_{RGBB} = 0.370 + 0.133$[Fe/H]. We then took the subsample of 7 metal-rich globular clusters (NGC104, 5927, 6356, 6624, 6569, 6637, and 6760) that are well-sampled, do not suffer from differential reddening and for which all or nearly all horizontal branch stars are RHBs rather than BHBs. The mean and sample deviation of their fractions are 31.7\% and 9.7\%. The correlation with metallicity is $f = (22.2 - 10.7$[Fe/H])\%. However there are only 7 points spanning a small range of metallicities relative to their extrapolation. Both sets of results are shown in Table \ref{table:PiottoClusters}. 

The result of low number density should not be surprising. If indeed the RGBB should be more easily measured in metal-rich populations, why was that of the Galactic bulge not measured decades ago? We argue that if the bulge RGBB had 20-30\% the number density of the RC it would have likely been detected, interpreted and integrated into the literature by this time. 

We suggest that the Galactic bulge has a higher helium to metallicity enrichment parameter (${\Delta}Y/{\Delta}Z$) than Galactic globular clusters, typically assumed to be $\sim$2. It is expected from stellar evolution models that helium enrichment reduces the lifetime of the RGBB.  \citet{2001ApJ...546L.109B} showed that for a 14 Gyr population, $R_{bump}$ should decrease by $\sim$10\%, for an increase in the initial stellar helium abundance ${\Delta}Y=0.06$.  More recently, \citet{2010ApJ...712..527D} estimated that for an [M/H]$=-0.35$ population, the lifetime of the RGBB will be decreased by one half as Y is increased from 0.26 to 0.31, and by two thirds when Y = 0.33. The numbers will be a little higher for a [M/H] $\approx 0$ population. An increased abundance of helium is also expected to lead to a longer lifetime for the horizontal branch \citep{1994A&A...285L...5R},  making $f_{Bump}$ an even further sensitive parameter to helium.  

\begin{figure}[H]
\begin{center}
\includegraphics[totalheight=0.54\textheight]{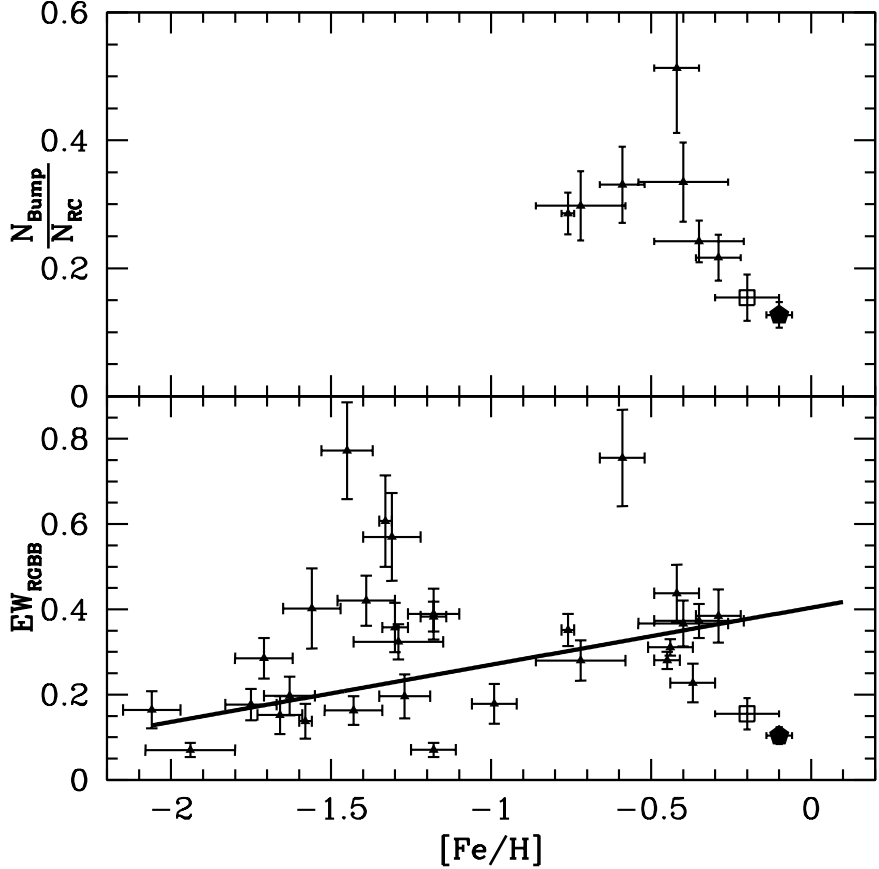}
\end{center}
\caption{TOP: The RGBB:RC ratio for 7 metal-rich globular clusters are shown as triangles, and that of the Galactic bulge is shown as the pentagon. BOTTOM: $EW_{RGBB}$ for 30 globular clusters are shown as triangles, and that of the Galactic bulge is shown as the pentagon. The solid black line is the theoretical relationship $EW_{RGBB} = 0.404 + 0.134$[Fe/H] for a 10 Gyr population. For both panels, the globular cluster metallicity is taken from \citet{2009A&A...508..695C}, and that of the Galactic bulge from \citet{2006ApJ...636..821F}. The data for the local dwarf spheroidal M32 is shown as the empty black square and is taken from \citet{2011ApJ...727...55M}, for which we assume  a metallicity error of 0.1 decs.}
\label{Fig:GCcalibration}
\end{figure}

\begin{figure}[H]
\begin{center}
\includegraphics[totalheight=0.7\textheight]{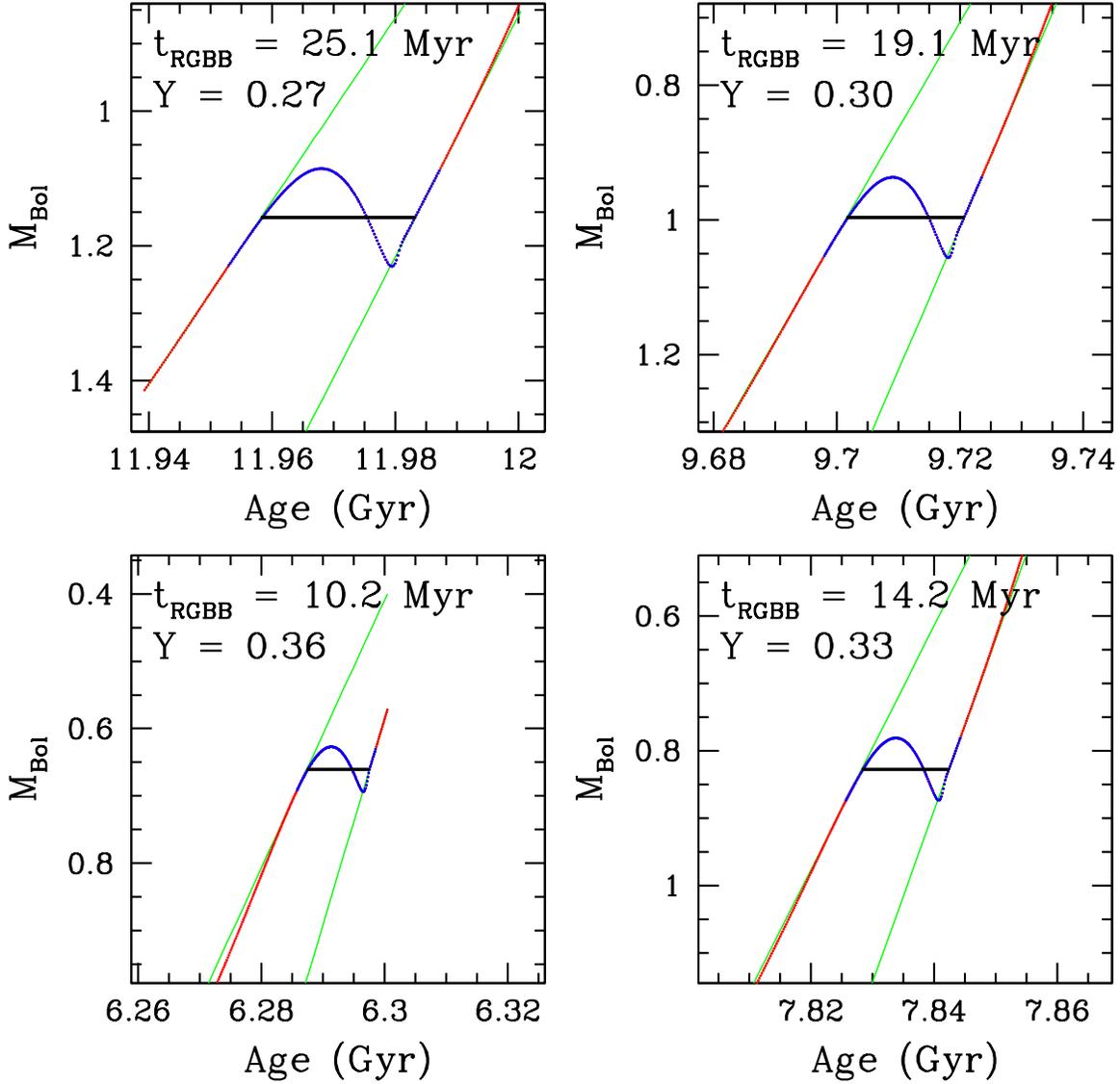}
\end{center}
\caption{The theoretical RGBB for 1 $M_{\odot}$ stars with solar metallicity, but with an initial helium content of 27,30,33 and 36\%, going clockwise from the top left. Blue dots represent the phase of RG evolution at the luminosity of the RGBB, and red dots the adjoining ascent of the red giant branch. The green lines are the tangent lines to the luminosity evolution before and after the RGBB phase, from which the lifetime is estimated. As the RG lifetime is very short compared to the total stellar lifetime, the luminosity evolution of a single star will be an excellent approximation to the luminosity function of a large number of stars with very small variations in initial mass.}
\label{Fig:StellarTracks}
\end{figure}

We have estimated the theoretical relationships using the YREC evolutionary code \citep{2010arXiv1005.0423D}. We assumed a primordial helium abundance of 25\% and a solar helium abundance of 27\% with a linear enrichment parameter ${\Delta}Y/{\Delta}Z$ in between, and we find $EW_{RGBB} = 0.404 + 0.134$[Fe/H]. For a [Fe/H]$=0$ star, the theoretical lifetime of the RGBB phase is ${t}_{RGBB} = 23.6 - 1.4(100Y-27)$ Myr, assuming a 10 Gyr population for both calculations. Since the lifetime of the RC is $\sim$100 Myr \citep{1997ApJ...486..201Y}, this implies an initial helium content of $\sim$35\% by mass for the typical metallicities of bulge stars. A higher helium abundance would be required to fit for the result $EW_{RGBB} = (0.104 \pm 0.020)$ mag, but that parameter is significantly more sensitive to age effects and disk contamination. We note that $EW_{RGBB}$ decreases with age, whereas $f_{bump}$ increases with age. The theoretical relation at [Fe/H]$=0$ and t$=10$ Gyr is  $EW_{RGBB} \approx 0.408 - 0.024(100Y-27) - 0.008(t-10\rm{Gyr})$ mag, requiring an initial helium content of 39\%. We report 35\% as our initial estimate, but we recognize that a more concrete estimate must account for the impact of disk contamination as well as integrating over a well-motivated composite population in metallicity and helium enrichment, tasks beyond the scope of this work. The effect of helium enrichment on the lifetime of the RGBB phase for a 1 $M_{\odot}$ star with solar metallicity is shown in Figure \ref{Fig:StellarTracks}.

\subsection{The Dispersion Parameter ${\Delta}{\sigma}^2$}
The best-fit dispersion in brightness of the RGBB stars is lower than that of the RC stars. For our canonical field, we find ${\sigma}_{RC} = 0.253 \pm 0.012$, and ${\sigma}_{RGBB} = 0.203 \pm 0.003$, implying a difference in dispersion ${\Delta}{\sigma}^2 = 0.05$, or ${\Delta}{\sigma} = 0.15$. This is surprising as the luminosity of the RGBB is a steeper function of metallicity than that of the RC. 

We estimate the expected intrinsic width of the RGBB as follows. We first take 195 RG+RC [Fe/H] measurements toward Baade's window from \citet{2008A&A...486..177Z}, which we then convert to [M/H] using the same prescription as \citet{2010arXiv1011.0457B}:  [$\alpha$/Fe]$=+0.30$ for [Fe/H] $\leq 0.0$, [$\alpha$/Fe]$=0.0$ for [Fe/H] $\geq 0.5$, and interpolating linearly in between. We then combine the results of \citet{2004AJ....127..958R} and \citet{2010ApJ...712..527D}, getting ${\Delta}I^{RGBB}_{RC} = 0.47 + 0.53$[M/H], assuming ${\Delta}I^{RGBB}_{RC} = {\Delta}V^{RGBB}_{ZAHB}$ as before and thereby allowing us to empirically estimate the brightness of each metallicity population. However, the brightness distribution is not just a linear transform of the metallicity distribution, as more metal-rich RGBB stars go through that phase earlier in their RG ascent, and thus the RGBB distribution will be biased toward the more metal-rich subset of stars. This effect is the same reason why the RGBB is harder to detect and classify in metal-poor clusters, in which the RGBB is brighter. We estimate the size of this bias by weighting each metallicity point by the observed relative number density of the Galactic bulge RG branch at that brightness, obtaining the following characteristics of RGBB stars:
\begin{equation}
 I_{i} = 0.47 + 0.53[M/H]
\end{equation}
\begin{equation}
 W_{i} = \exp\biggl[B*I_{i}\biggl],
\end{equation}

where $I_{i}$ is the estimated clump-centric brightness of the RGBB for that metallicity,  $W_{i}$ is the weight of that brightness component in the estimated distribution, and $B$ is the measured exponential slope of the RG branch, $0.547$. We calculate the weighted mean and variance:
\begin{equation}
 \overline{I}  = \frac{\sum_{1}^{N} W_{i} I_{i}}{\sum_{1}^{N} W_{i}}
\end{equation}
\begin{equation}
{\sigma_{I}}^2  = \frac{\sum_{1}^{N} W_{i} (I_{i} - \overline{I})^2}{\sum_{1}^{N} W_{i}},
\end{equation}
obtaining a mean $\overline{I}=0.560$ and ${\sigma_{I}}=0.157$. This is reassuringly lower than the measured dispersion ${\sigma}_{RGBB} = 0.203$ mag, a ``requirement'' as there will be additional variation induced by differential reddening, binary blending contamination and distance variation due to the large physical size of the bulge - contributions expected to similarly affect the RC. We briefly note the difference in the mean compared to our measurement 
could also be removed by discarding the small sample of expected RGBB (less than 5\%) that are within 0.25 mag of the RC as our fitting routine is not sensitive to those RGBB stars -- the mode of our expected distribution is at $\sim$0.71 mag. 

A smaller brightness dispersion of the RGBB than for the RC is a surprise. Performing the same calculation as above for RC stars, assuming lifetimes independent of metallicity and a brightness relation ${\Delta}I_{RC}/{\Delta}$[Fe/H] $= 0.14$ mag/dec \citep{2010AJ....140.1038P} implies a dispersion of ${\sigma}_{I} \approx 0.06$ mag. Galactic bulge RG stars have been measured to have similar proper motion distributions as their RC counterparts \citep{2007MNRAS.378.1165R}, an observational constraint that negates the possibility of a substantial bias-inducing age-metallicity-geometry correlation. There could be a bias induced in the parameters if the bulge RC has a skewed distribution as it is fit to a Gaussian, the residuals would then affect the fit to the RGBB. We do not expect this to be the case as the brightness distribution of the RC should be made very symmetric by geometrical dispersion. Forcing the intrinsic dispersion of the RGBB to be 0.05 mag higher than that of the RC increases ${\chi}^2$ by $\sim$320, demonstrating that a significantly different functional form would need to be assumed and justified to obtain a larger dispersion for the RGBB. We therefore conclude that the narrow luminosity function for the RGBB relative to the RC may be real. This could be achieved if there are subpopulations in the bulge that contribute substantially to the RC population and thus its dispersion but very weakly to the RGBB, such as a much more helium-rich population. In this scenario, it is the RC dispersion is too high rather than the RGBB dispersion being too low. 

\section{The Asymptotic Giant Branch Bump?}
The AGBB \citep{1998ApJ...495L..43G} may be present as a second peak in our magnitude distribution, at a brightness $\sim$1.06 mag higher than that of the RC, visible in the bottom panel of Figure \ref{Fig:markovscriptPaper}. Since the AGBB is a weak signal, it is necessary to constrain a third Gaussian in order to properly fit it, and we first do this by applying a ${\chi}^2$ penalty to solutions with a peak AGBB brightness outside the interval $1.00 <I_{RC} - I_{RGBB} < 1.30$. The best-fit then has a ${\Delta}{\chi}^2 \sim 125$, with a peak-brightness $\sim$1.12 mag brighter than the RC and a population of 0.8\% that of the RC. However, it has a seemingly unphysical dispersion of $\sim$0.11 mag, which is substantially narrower than that of the RC and the RGBB. This could be the case if the AGBB probes a much narrower range of stellar evolution than the RC or RGBB or if it is more sensitivity to an age-geometry-metallicity correlation, but we suspect it is like due to systematics in the fit. Constraining the AGBB to have the same dispersion as the RC yields a solution where the AGBB is $\sim$1.04 mag brighter than the RC in $I$, has a number density that is 1.8\% that of the RC, with a ${\Delta}{\chi}^2 \sim 79$. Constraining it to have the same dispersion as the RGBB yields an AGBB to RC ratio of $\sim$1.5\%, a separation in brightness of $\sim$1.08 mag, and ${\Delta}{\chi}^2 \sim 100$. The signal for the AGBB is not due to the RC of a foreground spiral arm or Galactic ring as the separation in brightness is independent of longitude, ruling out different geometrical configurations for the two peaks. It also shows up in the double-clump fields as seen in Figure \ref{Fig:DoubleBump}. This feature could provide an important insight into Galactic bulge evolution in the future should theoretical investigations and empirical calibrations better constrain its detailed properties. We note that the Galactic bulge AGBB:RC ratio of $(1.5\pm0.5)$\% is substantially lower than the $(6.0\pm1.8)$\% detected in M32 \citep{2011ApJ...727...55M}, which is perhaps surprising given that their RGBBs behave in a similar manner.

\section{Discussion and Conclusion}
\label{sec:Discussion}
We have demonstrated that the Galactic bulge RGBB peaks at or near the expected luminosity, but that its number-density is several times lower than would be expected from globular cluster calibrations. The brightness dispersion of the RGBB is lower than that of the RC. Both effects can be explained if there is a substantial, helium-rich population in the bulge that contributes to the RC but not the RGBB. 

This work will be difficult to extend to most of the Galactic bulge  as many sightlines have much higher brightness dispersions due to geometrical effects, leading to degeneracies in the fit to the RC and the RGBB. However, there are several promising avenues for further investigation. With our dataset, investigations of the AGBB and of the double clump-bump pairs toward what are probably the X-wings of the Milky Way bulge are both possible. Additionally, any large photometric dataset toward the inner few square degrees will be probing sightlines with minimal geometric dispersion, allowing a clearer separation of the RC and the RGBB. These are also the densest stellar fields, and so it might be possible to measure gradients. 

We find one literature mention that the Galactic bulge may be helium-rich. \citet{1994A&A...285L...5R} estimated $2 \leq {\Delta}Y/{\Delta}Z \leq 3$, for a helium content $0.30 \leq$ Y $\leq 0.35$. His analysis was based on the high value of the number ratio $R_{C}$ of RC to RG stars, an effect previously argued to be due to a relatively young stellar population \citep{1994AJ....107.2060P}. We comment on one chemical-enrichment model of interest. In their analysis of the split horizontal branch of the metal-rich globular cluster  Terzan 5, \citet{2010ApJ...715L..63D} showed that the helium-rich ejecta of massive asymptotic giant branch stars could lead to a substantially helium-enriched (${\Delta}Y \sim 0.07$) secondary population within $\sim$ 100 Myr. This is shorter than the currently expected $\sim$0.5 Gyr timescale of Galactic bulge formation \citep{2010arXiv1010.1469C,2010arXiv1011.0457B}. As our method generalizes well to any large stellar system, we wonder if a similarly low RGBB population density is to be found in the pseudobulges of nearby massive spiral galaxies, and if enhanced helium-enrichment is a characteristic of such populations.  

Enhanced helium enrichment, if confirmed, will lead to a downward revision of the 10+ Gyr age of the Galactic bulge determined from photometric isochrone fitting \citep{2003A&A...399..931Z,2010arXiv1011.0457B}. This would possibly bring it in line with the substantially younger ages detected via microlensing spectroscopy of dwarf stars \citep{2010A&A...512A..41B} and star counts of infrared carbon stars \citep{2002ApJ...574L..43C}.

\begin{table*}[ht]
\begin{center}
\caption{Measured globular cluster population parameters for the HST sample of \citet{2002A&A...391..945P}. [Fe/H] is taken from the globular cluster metallicity scale of \citet{2009A&A...508..695C}.}
\scalebox{0.95}
{\begin{tabular}{ l l l l l l}
\\
\hline\hline
Object & $N_{RGBB}$ & $N_{RC}$ & $EW_{RGBB}$ & [Fe/H] & ${\sigma}_{[Fe/H]}$ \\
\hline
NGC 104 & 87 & 306 & 0.35 & -0.76 & 0.02 \\ 
NGC 362 & 38 & - & 0.36 & -1.30 & 0.04 \\ 
NGC 1261 & 14 & - & 0.20 & -1.27 & 0.08 \\ 
NGC 1851 & 42 & - & 0.39 & -1.18 & 0.08 \\ 
NGC 1904 & 11 & - & 0.14 & -1.58 & 0.02 \\ 
NGC 2808 & 121 & - & 0.38 & -1.18 & 0.04 \\ 
NGC 5024 & 14 & - & 0.16 & -2.06 & 0.09 \\ 
NGC 5824 & 18 & - & 0.07 & -1.94 & 0.14 \\ 
NGC 5904 & 32 & - & 0.61 & -1.33 & 0.02 \\ 
NGC 5927 & 38 & 176 & 0.38 & -0.29 & 0.07 \\ 
NGC 5986 & 19 & - & 0.20 & -1.63 & 0.08 \\ 
NGC 6093 & 23 & - & 0.18 & -1.75 & 0.08 \\ 
NGC 6139 & 36 & - & 0.29 & -1.71 & 0.09 \\ 
NGC 6229 & 24 & - & 0.16 & -1.43 & 0.09 \\ 
NGC 6266 & 18 & - & 0.07 & -1.18 & 0.07 \\ 
NGC 6284 & 31 & - & 0.57 & -1.31 & 0.09 \\ 
NGC 6356 & 88 & 364 & 0.37 & -0.35 & 0.14 \\ 
NGC 6388 & 188 & - & 0.28 & -0.45 & 0.04 \\ 
NGC 6402 & 51 & - & 0.42 & -1.39 & 0.09 \\ 
NGC 6441 & 257 & - & 0.31 & -0.44 & 0.07 \\ 
NGC 6522 & 46 & - & 0.77 & -1.45 & 0.08 \\ 
NGC 6569 & 35 & 118 & 0.28 & -0.72 & 0.14 \\ 
NGC 6624 & 42 &  83 & 0.44 & -0.42 & 0.07 \\ 
NGC 6637 & 45 & 135 & 0.76 & -0.59 & 0.07 \\ 
NGC 6638 & 15 & - & 0.18 & -0.99 & 0.07 \\ 
NGC 6760 & 47 & 139 & 0.37 & -0.40 & 0.14 \\ 
NGC 6864 & 62 & - & 0.32 & -1.29 & 0.14 \\ 
NGC 6934 & 18 & - & 0.40 & -1.56 & 0.09 \\ 
NGC 7089 & 12 & - & 0.15 & -1.66 & 0.07 \\ 
NGC 6304 & 25 & - & 0.23 & -0.37 & 0.07 \\ 
\hline
\end{tabular}
\label{table:PiottoClusters}
}\end{center}
\end{table*}

\acknowledgments
DMN and AG were partially supported by the NSF grant AST-0757888. The OGLE project has received funding from the European Research Council
under the European Community's Seventh Framework Programme
(FP7/2007-2013) / ERC grant agreement no. 246678 to AU. We thank P.B. Stetson, R.A. Benjamin and P. Fouqu\'e  for informative discussions. We also thank the anonymous referee for his suggested adjustments.


\begin{thebibliography}{}
\bibitem[Babusiaux 
\& Gilmore(2005)]{2005MNRAS.358.1309B} Babusiaux, C., \& Gilmore, G.\ 2005, \mnras, 358, 1309 
\bibitem[Babusiaux et 
al.(2010)]{2010A&A...519A..77B} Babusiaux, C., et al.\ 2010, \aap, 519, A77 
\bibitem[Bellazzini et al.(2002)]{2002AJ....124.3222B} Bellazzini, M., 
Ferraro, F.~R., Origlia, L., Pancino, E., Monaco, L., 
\& Oliva, E.\ 2002, \aj, 124, 3222 
\bibitem[Bensby et 
al.(2010)]{2010A&A...512A..41B} Bensby, T., et al.\ 2010, \aap, 512, A41 
\bibitem[Bono et al.(2001)]{2001ApJ...546L.109B} Bono, G., Cassisi, S., 
Zoccali, M., \& Piotto, G.\ 2001, \apjl, 546, L109 
\bibitem[Bono et 
al.(2007)]{2007A&A...476..779B} Bono, G., Caputo, F., \& Di Criscienzo, M.\ 2007, \aap, 476, 779 
\bibitem[Brown et al.(2010)]{2010arXiv1011.0457B} Brown, T.~M., et al.\ 
2010, arXiv:1011.0457 
\bibitem[Cabrera-Lavers et 
al.(2008)]{2008A&A...491..781C} Cabrera-Lavers, A., Gonz{\'a}lez-Fern{\'a}ndez, C., Garz{\'o}n, F., Hammersley, P.~L., \& L{\'o}pez-Corredoira, M.\ 2008, \aap, 491, 781 
\bibitem[Carretta et 
al.(2009)]{2009A&A...508..695C} Carretta, E., Bragaglia, A., Gratton, R., D'Orazi, V., \& Lucatello, S.\ 2009, \aap, 508, 695 
\bibitem[Cassisi 
\& Salaris(1997)]{1997MNRAS.285..593C} Cassisi, S., \& Salaris, M.\ 1997, \mnras, 285, 593 
\bibitem[Cescutti 
\& Matteucci(2010)]{2010arXiv1010.1469C} Cescutti, G., \& Matteucci, F.\ 2010, arXiv:1010.1469 
\bibitem[Cole 
\& Weinberg(2002)]{2002ApJ...574L..43C} Cole, A.~A., \& Weinberg, M.~D.\ 2002, \apjl, 574, L43 
\bibitem[D'Antona et al.(2010)]{2010ApJ...715L..63D} D'Antona, F., Ventura, 
P., Caloi, V., D'Ercole, A., Vesperini, E., Carini, R., 
\& Di Criscienzo, M.\ 2010, \apjl, 715, L63 
\bibitem[Delahaye et al.(2010)]{2010arXiv1005.0423D} Delahaye, F., 
Pinsonneault, M.~H., Pinsonneault, L., 
\& Zeippen, C.~J.\ 2010, arXiv:1005.0423 
\bibitem[Di Cecco et al.(2010)]{2010ApJ...712..527D} Di Cecco, A., et al.\ 
2010, \apj, 712, 527 
\bibitem[Ferraro et al.(2009)]{2009Natur.462..483F} Ferraro, F.~R., et al.\ 
2009, \nat, 462, 483 
\bibitem[Fulbright et al.(2006)]{2006ApJ...636..821F} Fulbright, J.~P., 
McWilliam, A., \& Rich, R.~M.\ 2006, \apj, 636, 821 
\bibitem[Fulbright et al.(2007)]{2007ApJ...661.1152F} Fulbright, J.~P., 
McWilliam, A., \& Rich, R.~M.\ 2007, \apj, 661, 1152 
\bibitem[Gallart(1998)]{1998ApJ...495L..43G} Gallart, C.\ 1998, \apjl, 495, 
L43 
\bibitem[Girardi 
\& Salaris(2001)]{2001MNRAS.323..109G} Girardi, L., \& Salaris, M.\ 2001, \mnras, 323, 109 
\bibitem[Holland et al.(1996)]{1996AJ....112.1035H} Holland, S., Fahlman, 
G.~G., \& Richer, H.~B.\ 1996, \aj, 112, 1035 
\bibitem[Lee et al.(2003)]{2003AJ....126.2840L} Lee, M.~G., et al.\ 2003, 
\aj, 126, 2840 
\bibitem[Majewski et al.(1999)]{1999ApJ...520L..33M} Majewski, S.~R., 
Siegel, M.~H., Patterson, R.~J., \& Rood, R.~T.\ 1999, \apjl, 520, L33 
\bibitem[McWilliam 
\& Zoccali(2010)]{2010ApJ...724.1491M} McWilliam, A., \& Zoccali, M.\ 2010, \apj, 724, 1491 
%
\bibitem[Monachesi et al.(2011)]{2011ApJ...727...55M} Monachesi, A., 
Trager, S.~C., Lauer, T.~R., Freedman, W., Dressler, A., Grillmair, C., 
\& Mighell, K.~J.\ 2011, \apj, 727, 55 
\bibitem[Monaco et al.(2002)]{2002ApJ...578L..47M} Monaco, L., Ferraro, 
F.~R., Bellazzini, M., \& Pancino, E.\ 2002, \apjl, 578, L47 
\bibitem[Monelli et al.(2010)]{2010ApJ...718..707M} Monelli, M., Cassisi, 
S., Bernard, E.~J., Hidalgo, S.~L., Aparicio, A., Gallart, C., 
\& Skillman, E.~D.\ 2010, \apj, 718, 707 
\bibitem[Nataf et al.(2010)]{2010ApJ...721L..28N} Nataf, D.~M., Udalski, 
A., Gould, A., Fouqu{\'e}, P., \& Stanek, K.~Z.\ 2010, \apjl, 721, L28 
\bibitem[Nishiyama et al.(2005)]{2005ApJ...621L.105N} Nishiyama, S., et 
al.\ 2005, \apjl, 621, L105 
\bibitem[Nishiyama et al.(2009)]{2009ApJ...696.1407N} Nishiyama, S., 
Tamura, M., Hatano, H., Kato, D., Tanab{\'e}, T., Sugitani, K., 
\& Nagata, T.\ 2009, \apj, 696, 1407 
\bibitem[Ortolani et 
al.(2001)]{2001A&A...376..878O} Ortolani, S., Barbuy, B., Bica, E., Renzini, A., Zoccali, M., Rich, R.~M., \& Cassisi, S.\ 2001, \aap, 376, 878
\bibitem[Paczynski et al.(1994)]{1994AJ....107.2060P} Paczynski, B., 
Stanek, K.~Z., Udalski, A., Szymanski, M., Kaluzny, J., Kubiak, M., 
\& Mateo, M.\ 1994, \aj, 107, 2060 
 \bibitem[Pietrzy{\'n}ski et al.(2010)]{2010AJ....140.1038P} 
Pietrzy{\'n}ski, G., G{\'o}rski, M., Gieren, W., Laney, D., Udalski, A., 
\& Ciechanowska, A.\ 2010, \aj, 140, 1038 
\bibitem[Piotto et 
al.(2002)]{2002A&A...391..945P} Piotto, G., et al.\ 2002, \aap, 391, 945 
\bibitem[Rattenbury et al.(2007a)]{2007MNRAS.378.1064R} Rattenbury, N.~J., 
Mao, S., Sumi, T., \& Smith, M.~C.\ 2007, \mnras, 378, 1064 
\bibitem[Rattenbury et al.(2007b)]{2007MNRAS.378.1165R} Rattenbury, N.~J., 
Mao, S., Debattista, V.~P., Sumi, T., Gerhard, O., 
\& de Lorenzi, F.\ 2007, \mnras, 378, 1165 
\bibitem[Renzini(1994)]{1994A&A...285L...5R} Renzini, A.\ 1994, \aap, 285, L5 
\bibitem[Rey et al.(2004)]{2004AJ....127..958R} Rey, S.-C., Lee, Y.-W., 
Ree, C.~H., Joo, J.-M., Sohn, Y.-J., \& Walker, A.~R.\ 2004, \aj, 127, 958 
\bibitem[Riello et 
al.(2003)]{2003A&A...410..553R} Riello, M., et al.\ 2003, \aap, 410, 553 
\bibitem[Salaris et al.(1993)]{1993ApJ...414..580S} Salaris, M., Chieffi, 
A., \& Straniero, O.\ 1993, \apj, 414, 580 
\bibitem[Stanek(1996)]{1996ApJ...460L..37S} Stanek, K.~Z.\ 1996, \apjl, 
460, L37 
\bibitem[Stanek et al.(1997)]{1997ApJ...477..163S} Stanek, K.~Z., Udalski, 
A., Szymanski, M., Kaluzny, J., Kubiak, M., Mateo, M., 
\& Krzeminski, W.\ 1997, \apj, 477, 163 
\bibitem[Sumi(2004)]{2004MNRAS.349..193S} Sumi, T.\ 2004, \mnras, 349, 193 
\bibitem[Udalski(2003)]{2003AcA....53..291U} Udalski, A.\ 2003, Acta 
Astronomica, 53, 291 
\bibitem[Udalski(2003)]{2003ApJ...590..284U} Udalski, A.\ 2003, \apj, 590, 
284 
\bibitem[Udalski et al.(2008)]{2008AcA....58...69U} Udalski, A., Szymanski, 
M.~K., Soszynski, I., \& Poleski, R.\ 2008, Acta Astronomica, 58, 69 
\bibitem[Yi et al.(1997)]{1997ApJ...486..201Y} Yi, S., Demarque, P., 
\& Oemler, A., Jr.\ 1997, \apj, 486, 201 
\bibitem[Zinn 
\& West(1984)]{1984ApJS...55...45Z} Zinn, R., \& West, M.~J.\ 1984, \apjs, 55, 45 
\bibitem[Zoccali et al.(1999)]{1999ApJ...518L..49Z} Zoccali, M., Cassisi, 
S., Piotto, G., Bono, G., \& Salaris, M.\ 1999, \apjl, 518, L49 
\bibitem[Zoccali et 
al.(2003)]{2003A&A...399..931Z} Zoccali, M., et al.\ 2003, \aap, 399, 931 
\bibitem[Zoccali et 
al.(2008)]{2008A&A...486..177Z} Zoccali, M., Hill, V., Lecureur, A., Barbuy, B., Renzini, A., Minniti, D., G{\'o}mez, A., \& Ortolani, S.\ 2008, \aap, 486, 177 
\end{thebibliography}
\end{document}